\pdfoutput=1
\documentclass[a4paper]{article}

\usepackage{itgspeech2025}    
\usepackage{times}            
\usepackage[english]{babel}   
\usepackage[ansinew]{inputenc}
\usepackage[T1]{fontenc}      
\usepackage[sort&compress,numbers]{natbib}	
\usepackage{amsmath,amssymb}
\usepackage{graphicx}
\usepackage[colorlinks=false,pdfborder={0 0 0}]{hyperref}
\usepackage{booktabs}
\usepackage{multirow}
\usepackage{cleveref}
\usepackage{balance}
\usepackage[separate-uncertainty=true]{siunitx}
\usepackage{subcaption}
\usepackage{tikz}
\usepackage{csquotes}
\usepackage[acronym, shortcuts, toc,nonumberlist]{glossaries}
\usetikzlibrary{positioning}

\newcommand{\veryshortarrow}[1][4pt]{\mathrel{%
   \hbox{\rule[\dimexpr\fontdimen22\textfont2-.2pt\relax]{#1}{.4pt}}%
   \mkern-4mu\hbox{\usefont{U}{lasy}{m}{n}\symbol{41}}}}

\newcolumntype{H}{>{\setbox0=\hbox\bgroup}c<{\egroup}@{}}   

\newacronym{SGMSE}{SGMSE}{Score-based Generative Model for Speech Enhancement}
\newacronym{SNR}{SNR}{Signal-to-Noise Ratio}
\newacronym{WER}{WER}{Word Error Rate}
\newacronym{SRMR}{SRMR}{Speech-to-Reverberation Modulation Ratio}
\newacronym{RIR}{RIR}{room impulse response}
\newacronym{SDE}{SDE}{stochastic differential equation}
\newacronym{ASR}{ASR}{Automatic Speech Recognition}

\sloppypar

\title{On the Application of Diffusion Models for Simultaneous Denoising \\ and Dereverberation}
\author{Adrian Meise, Tobias Cord-Landwehr, Reinhold Haeb-Umbach}

\address{Paderborn University, Communications Engineering Department, Paderborn, Germany\\
  Email: \texttt{\{meise,cord,haeb\}@nt.upb.de}}

\begin{document}

\maketitle
\setlength{\textfloatsep}{5pt plus 0.0pt minus 2.0pt}
\setlength{\floatsep}{5pt plus 0.0pt minus 2.0pt}
\setlength{\intextsep}{5pt plus 0.0pt minus 2.0pt}

\begin{abstract}
Diffusion models have been shown to achieve natural-sounding  enhancement of speech degraded by noise or reverberation. However, their simultaneous denoising and dereverberation capability has so far not been studied much, although this is arguably the most common scenario in a practical application.
In this work, we investigate different approaches to enhance noisy and/or reverberant speech. We examine the cascaded application of models, each trained on only one of the distortions, and compare it with  a single model, trained either solely on data that is both noisy and reverberated, or trained on data comprising subsets of purely noisy, of purely reverberated, and of noisy reverberant speech. Tests are performed both on artificially generated and real recordings of noisy and/or reverberant data. 
The results show that, when using the cascade of models, satisfactory results are only achieved if they are applied in the order of the dominating distortion. 
If only a single model is desired that can operate on all distortion scenarios, the best compromise appears to be a model trained on the aforementioned three subsets of degraded speech data.
\end{abstract}

\section{Introduction}
Speech enhancement aims to remove environmental disturbances from a recorded speech signal. While there can be a huge variety of distortions, the literature in the field mostly concentrates on the removal of noise and reverberation.  

Earlier systems were based on model-based approaches, typically employing statistical models, but for several years already, they have been mostly replaced by  hybrid techniques that blend model-based and data-driven components or by purely data-driven methods \cite{23_ochieng_se_overview,24_haeb_se_overview}. Data-driven speech enhancement is based on deep neural networks. The plethora of models that have been proposed can be broadly categorized as being of discriminative or generative nature. Discriminative approaches compute a point estimate, such as the mean of the clean speech posterior, whereas generative  models aim at sampling from the posterior. 


Recently, speech enhancement based on diffusion models, which fall in the class of generative models, has been reported to achieve excellent results \cite{22_Lu_diffusion_se, welker_22_sgmse, 23_Gonzalez_heun_binaural_se, 24_Jukic_schroedinger_diff}.   
A popular choice are score-based diffusion models, such as \cite{welker_22_sgmse}. Here, enhancement is performed by modeling a forward and reverse diffusion process, where the latter is solved with a trained score model.
These systems have been shown to produce very naturally-sounding enhanced speech, effectively removing noise or reverberation.
Furthermore, they can be easily adapted to new tasks such as bandwidth extension \cite{lemercier_23_restoration} and joint dereverberation and \gls{RIR} estimation \cite{lemercier_25_buddy}.

However, assessing their performance in domains encompassing both noise and reverberation, in particular on real recordings, has so far not been studied systematically. While \cite{22_Lu_diffusion_enhancement, welker_22_sgmse, richter_23_sgmse+, lemercier_23_storm, lemercier_23_restoration, scheibler24_interspeech}  address denoising and dereverberation individually, they do not further investigate the combination of both objectives.
In \cite{lemercier_23_posterior_sampling}, dereverberation is performed in the presence of noise, but the focus is on dereverberation while the noise is considered as a pass-through artifact that needs to be handled via a postprocessing stage.
Further, recent challenges like \cite{urgent_challenge} consider noise and reverberation only together with a variety of other distortion types.

The study in \cite{kimura_24_msgmse} considered \gls{SGMSE} for multi-channel joint dereverberation and denoising, which was further extended by the addition of multiple input data streams in \cite{nakatani_24_msgmse_multistream}, and multi-modal generation and enhancement in \cite{23_Yang_Usee_multimodal}. But these works focused on the multi-channel extension and multi-modal applicability, and  discussed the problem of joint dereverberation and enhancement only in passing. Furthermore, \cite{kimura_24_msgmse, nakatani_24_msgmse_multistream} considered only \glspl{SNR} in the range between \SIrange{10}{14}{\decibel} and only a small distance of \SIrange{0.5}{1.5}{\meter} between the speech source and the microphones, lacking a comprehensive evaluation of the denoising and dereverberation scenario in a realistic far-field scenario. A signal-level performance assessment of diffusion models on real recordings in a noisy reverberant environment appears to be completely missing so far. 

In this work, we systematically investigate how joint dereverberation and denoising can be best performed using diffusion models. We look at both cascading components, one specialized on denoising and one on dereverberation, and on using a single model for both tasks.
Experiments are carried out on a variety of data sets: data that is only degraded by additive noise (WSJ-CHiME3 \cite{richter_23_sgmse+} and EARS-WHAM \cite{richter_24_ears}), data that is only degraded by reverberation (WSJ-Reverb \cite{richter_23_sgmse+} and EARS-Reverb \cite{richter_24_ears}), and data that exhibits both distortion types (EARS-WHAMR!), including a data set of real recordings in a noisy reverberant environment (REVERB) \cite{reverb_13}.
Enhancement performance is measured with objective metrics that assess speech quality and intelligibility, and by measuring the word error rate of a downstream speech recognizer, which cannot be fooled by hallucinations, a common issue of diffusion models, that may not be caught by signal-level metrics.
The results show that the order in which models trained on a single task are cascaded for solving the joint problem is crucial for good speech enhancement performance. But the experiments also reveal that  in reverberant and low-SNR conditions, the cascade falls short of the performance of a single model, trained on both distortion types.


This work is structured as follows.
First, we give a short overview of diffusion-model-based speech enhancement in \Cref{sec:diffusion_enhancement}. Then, we discuss the theoretical motivation for the use of diffusion models for simultaneous denoising and dereverberation and present possible alternatives in terms of cascades of individual models. \Cref{sec:experiment_setup} describes the experimental setup before the results are presented and discussed in \Cref{sec:results}.

\section{Diffusion-based enhancement}
\label{sec:diffusion_enhancement}

We model the speech signal $y(\tilde{t})$  observed at the microphone at sampling time $\tilde{t}$ to be the superposition of the  target source signal $x(\tilde{t})$, which is reverberated by the \gls{RIR} $h(\tilde{t})$, and additive  $n(\tilde{t})$:
\begin{equation}
y(\tilde{t}) = x(\tilde{t}) \ast h(\tilde{t}) + n(\tilde{t}).
\end{equation}
Here, the goal of speech enhancement is the retrieval of the speech signal $x(\tilde{t})$, where the removal of $n(\tilde{t})$ is referred to as denoising  
and the reversal of the convolution by $h(\tilde{t})$ as dereverberation.
In the following, processing is carried out in the complex STFT domain, where the signal vectors $\mathbf{x}$, $\mathbf{y}$, $\mathbf{n}\in \mathbb{C}^{F*K}$ are flattened complex-valued vectors, where $F$ is the number of frequency bins and $K$ the number of time frames.

\subsection{Score-based generative models}
The  principle of score-based generative models for speech enhancement is based on a diffusion process described by the formalism of a \gls{SDE} as derived in \cite{welker_22_sgmse, richter_23_sgmse+}.

The forward process is modeled as
\begin{equation}
\mathrm{d}\mathbf{x}_t = \mathbf{f}(\mathbf{x}_t,\mathbf{y})\mathrm{d}t + g(t)\mathrm{d}\mathbf{w} .
\end{equation}
Here,  $\mathbf{f}$ is referred to as the drift term, while $g$ denotes the diffusion term and $\mathbf{w}$ describes a standard Wiener process. With increasing time $t \in [0,T]$ of the process, the data $\mathbf{x}$ is corrupted with an increasing amount of noise. By selecting the drift term as $\mathbf{f} = \gamma (\mathbf{y} - \mathbf{x}_t)$, a dependency on the observation signal $\mathbf{y}$ is introduced. Therefore, the clean speech sample is turned into a noisy one that incorporates information and characteristics of the observation, thus creating a relation between them as needed for speech enhancement.

The corresponding reverse \gls{SDE} can be derived as
\begin{equation}
 \mathrm{d}\mathbf{x}_t = [-\mathbf{f}(\mathbf{x}_t,\mathbf{y}) + g(t)^{2} \mathbf{s}_{\boldsymbol{\theta}}(\mathbf{x}_t, \mathbf{y},t)]\mathrm{d}t + g(t)\mathrm{d}\mathbf{\bar{w}},
\end{equation}
where $\mathbf{\bar{w}}$ denotes the Wiener process flowing in reverse direction.
Denoising score matching \cite{vincent_11_denoising_score_matching} aims at training a score model $\mathbf{s}_{\boldsymbol{\theta}}(\mathbf{x}_t,\mathbf{y},t)$ with learnable parameters $\boldsymbol{\theta}$ that predicts the score  $\nabla_{\mathbf{x}_t} \log p_t(\mathbf{x}_t|\mathbf{y})$.
This can be used to solve the reverse diffusion process with numerical methods as predictor-corrector samplers \cite{song_21_sgm}.

\subsection{Joint Denoising \& Dereverberation}
\label{sec:joint_denoising_dereverberation}
During the forward and reverse process, the model is tasked to learn the transfer from the tractable distribution described by the observed signal and the added artificial Gaussian noise at the end time $t=T$ of the forward diffusion process to the distribution of clean speech at process time $t=0$. 
This added noise is supposed to mask the characteristics of the disturbances present in the observed signals, such as the environmental noise. As a consequence, the same approach is not only able to perform denoising \cite{welker_22_sgmse}, but also dereverberation \cite{richter_23_sgmse+} or bandwidth extension \cite{lemercier_23_restoration}, since the individual disturbance type only influences the data distribution at time $T$ from which the clean sample is retrieved. The disturbance in the observation is then interpreted as noise that is removed during solving the reverse process, causing the independence of processing from the actual disturbance and allowing the application of the same procedure to arbitrary distortions.

Therefore, the model should also be able to enhance noisy and reverberant data as well, where both disturbance types can have different degrees of severity.  
Only the training data needs to be adapted while the principal approach remains the same. For this, there are two options: Either only noisy reverberant data can be presented to the model, or a mixture of only noisy, only reverberated data and - optionally - noisy reverberant data can be used.

\begin{figure}[t]
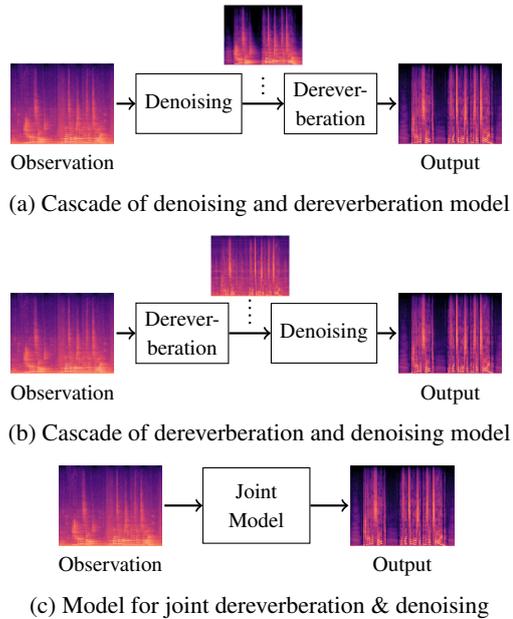

    \begin{subfigure}[t]{\columnwidth}
        \centering
        \input{block_diagram/block_diagram_1}
        \vspace{-0.25em}
        \caption{Cascade of denoising and dereverberation model}\vspace{0.5em}
    \end{subfigure}
    
    \begin{subfigure}[t]{\columnwidth}
        \centering
        \input{block_diagram/block_diagram_2}
        \vspace{-0.25em}
        \caption{Cascade of dereverberation and denoising model}\vspace{0.5em}
    \end{subfigure}

    \begin{subfigure}[t]{\columnwidth}
        \centering
        \input{block_diagram/block_diagram_3}
        \vspace{-0.25em}
        \caption{Model for joint dereverberation \& denoising}
    \end{subfigure}
    \caption{Overview of the possible approaches for combining denoising and dereverberation.}
    \label{fig:block_diagrams}
\end{figure}

An alternative to the training on the joint task is the cascade of two individual models, one denoising and one dereverberating model.
The justification for the cascade is the assumption  that the model for one distortion is transparent to the other, i.e., if the first model does denoising, it leaves reverberation unaffected, which is subsequently tackled by the second model.
Fig.~\ref{fig:block_diagrams} visualizes these alternative approaches for the enhancement of noisy and reverberant input data.

\section{Experimental Setup}
\label{sec:experiment_setup}
As a basis for all experiments, the SGMSE+ architecture from \cite{richter_23_sgmse+} is used, a score-based generative diffusion approach as described in \Cref{sec:diffusion_enhancement}. 
We train all models from scratch with the same parametrization as in \cite{richter_23_sgmse+, richter_24_ears} using the code from \cite{sgmse_repo}. 

We trained four model configurations: a denoising network trained on noisy data (\enquote{Denoise}), a dereverberation network trained on reverberated data (\enquote{Dereverb}), a joint network trained on reverberated data to which noise is added (\enquote{Joint Model}), and a multi-objective network, which is trained on all three types of data, with one third from each (\enquote{Mixed Objective}).
It is important to note that the amount of training data is the same for all configurations. 

The models are trained on and applied to scenarios based on either the WSJ corpus \cite{wsj} or the EARS corpus \cite{richter_24_ears}. 
On WSJ, for denoising and dereverberation, the synthetically created datasets WSJ-CHiME3 and \mbox{WSJ-Reverb}, as used in \cite{richter_23_sgmse+}, are employed respectively. 
For joint denoising and dereverberation, the real test set of the REVERB challenge \cite{reverb_13} is used, which consists of rerecordings of the  WSJ0 prompts  with a \gls{SNR} of $\approx$\SI{20}{\decibel} and an average $T_{60}$ time of \SI{0.7}{\s}. 

On the EARS corpus, the previously introduced EARS-WHAM and EARS-Reverb datasets from \cite{richter_24_ears} are used as denoising and dereverberation datasets. 
For evaluating joint dereverberation and denoising, we simulate a new reverberant and noisy variant of EARS, which we call EARS-WHAMR!. Here, we add the noise samples used for EARS-WHAM to the reverberant data of EARS-Reverb using the same parameterization as for the individual datasets.

The two employed base corpora, WSJ and EARS, have quite different properties. For WSJ, the source signals show low variety. Additionally, the REVERB data exhibits a high SNR, resulting in the reverberation being the dominant distortion that is to be removed during enhancement.
For EARS, the source signals  exhibit a broader dynamic range, including emotional and highly dynamic speech.
In addition, for EARS-WHAMR!, both reverberation and noise have a severe impact in terms of distortion levels. 
All models are evaluated in a matched scenario, i.e., models evaluated on WSJ-based data are trained using only the respective WSJ training datasets and vice versa for EARS.

Model performance is evaluated with the DNSMOS score \cite{dnsmos_21} for speech quality assessment in order to emulate human listening perception, with the \gls{SRMR} \cite{srmr} as a measure of the degree of dereverberation, and the \gls{WER} as an indicator for the speech (machine) intelligibility. 
For the synthetic test sets, we additionally report ESTOI \cite{estoi}.
Since we evaluate both on reverberant and anechoic data, we do not use the widely used Si-SDR \cite{roux_19_sisdr} as it is ill-defined for reverberant conditions \cite{drude2019sms}.
The QuartzNet-based model from the NeMo toolkit \cite{nemo} is used as \gls{ASR} engine, same as in \cite{richter_24_ears}. That model has not been finetuned on noisy or reverberated data, and is thus susceptible to the distortions and artifacts introduced by speech enhancement.
While a single metric alone may not capture all aspects, the combined assessment with the above set of metrics gives a more complete picture of the enhancement performance. For instance, a high WER but a high DNSMOS indicates a natural-sounding model output with speech-like artifacts that no longer carry content, possibly indicating  hallucinations.

\section{Results}
\label{sec:results}
For the evaluation, we first examine the speech enhancement performance for environments with a single disturbing source and compare the difference between the models trained on a single disturbance to the joint training and model cascades. 
After that, the individual models are applied to data that is both noisy and reverberant.

\begin{table*}[h!]
    \centering
    \caption{System comparison for simultaneous dereverberation and denoising on the (real) REVERB challenge and (synthetic) EARS-WHAMR! sets. All models are trained in a matched scenario.}
        \sisetup{table-format=1.2+-1.2}
 \newcolumntype{H}{>{\setbox0=\hbox\bgroup}c<{\egroup}@{}}   
\sisetup{
text-series-to-math = true,
propagate-math-font = true
}
    \begin{tabular}{l c c H c || c c c c}
    \toprule
         \multirow{2}{*}{Model} & \multicolumn{4}{c}{REVERB} & \multicolumn{4}{c}{EARS-WHAMR!} \\
         \cmidrule(lr){2-5}\cmidrule(lr){6-9}
         & {SRMR} & {DNSMOS} & {PESQ} & {WER} & {SRMR} & {DNSMOS} & {ESTOI} & {WER} \\ 
         \midrule
         Observation & \num{3.16} \footnotesize{$\pm$ \num{0.95}} & \num{2.82} \footnotesize{$\pm$ \num{0.28}} & -- & \num{21.2}
         &  \hphantom{1}\num{2.76} \footnotesize{$\pm$ \num{1.32}} &  \num{2.46} \footnotesize{$\pm$ \num{0.22}} &  \num{0.27} \footnotesize{$\pm$ \num{0.11}} & \num{68.5}\\
         \midrule
         Denoise  & \num{7.62} \footnotesize{$\pm$ \num{3.32}} &  \num{3.70} \footnotesize{$\pm$ \num{0.31}} &  -- & \num{59.4} & $\mathbf{15.05}$ \footnotesize{$\pm$ \num{9.76}} & \num{2.86} \footnotesize{$\pm$ \num{0.54}} & \num{0.28} \footnotesize{$\pm$ \num{0.17}} & \num{86.5} \\
         Dereverberation & \num{5.54} \footnotesize{$\pm$ \num{2.00}} &  \num{3.10} \footnotesize{$\pm$ \num{0.34}} &  -- & $\mathbf{13.5}$ & \hphantom{1}\num{3.26} \footnotesize{$\pm$ \num{1.81}} & \num{2.59} \footnotesize{$\pm$ \num{0.26}} & \num{0.30} \footnotesize{$\pm$ \num{0.13}} & \num{72.3} \\
        \midrule
        Denoise $\rightarrow$ Dereverb  & $\mathbf{7.76}$ \footnotesize{$\pm$ \num{3.52}} &   \num{3.70} \footnotesize{$\pm$ \num{0.30}} &  -- & \num{58.8} & \num{14.89} \footnotesize{$\pm$ \num{9.89}} & \num{2.82} \footnotesize{$\pm$ \num{0.53}} & \num{0.28} \footnotesize{$\pm$ \num{0.17}} & \num{86.7} \\
        Dereverb $\rightarrow$ Denoise & \num{7.63} \footnotesize{$\pm$ \num{2.93}} &  \num{3.83} \footnotesize{$\pm$ \num{0.24}} &  -- & \num{21.5} & \hphantom{1}\num{8.73} \footnotesize{$\pm$ \num{6.41}} & \num{2.87} \footnotesize{$\pm$ \num{0.55}} & \num{0.29} \footnotesize{$\pm$ \num{0.20}} & \num{80.8}\\
        Joint Model & \num{7.48} \footnotesize{$\pm$ \num{2.91}} &  \num{3.86} \footnotesize{$\pm$ \num{0.24}} &  -- & \num{17.2} & \hphantom{1}\num{9.01} \footnotesize{$\pm$ \num{4.35}} & $\mathbf{3.84}$ \footnotesize{$\pm$ \num{0.30}} & $\mathbf{0.58}$ \footnotesize{$\pm$ \num{0.16}} & $\mathbf{42.6}$\\
        Mixed Objective & \num{7.30} \footnotesize{$\pm$ \num{2.94}} &  $\mathbf{3.91}$ \footnotesize{$\pm$ \num{0.21}} &  -- & \num{19.0} & \hphantom{1}\num{8.75} \footnotesize{$\pm$ \num{4.24}} & \num{3.80} \footnotesize{$\pm$ \num{0.31}} & \num{0.56} \footnotesize{$\pm$ \num{0.17}} & \num{44.2}\\
        \bottomrule
    \end{tabular}
    \label{tab:joint_results}
\end{table*}

\subsection{Performance in environments with a single disturbance}
\Cref{tab:results_denoise} shows the denoising performance of the models.
On WSJ-CHiME3 (\Cref{tab:wsj_noise}), it can be seen that no large performance differences occur among the models.
In terms of \gls{WER}, the dedicated denoising network performs best as expected. Pre- or post-processing by an additional dereverberation network only introduces little artefacts and provides similar results as the mono-objective denoising model. The joint model, trained on noisy reverberant data, shows some degradation, which is to be expected due to the train-test mismatch. This degradation can be mitigated by also showing only noisy or reverberant data to the model during training, as done in the \textit{Mixed} setup.

For the more challenging EARS-WHAM set (\Cref{tab:ears_noise}), significant changes are apparent. Here, the dedicated denoising network still performs best. However, as soon as the denoising model is cascaded with a dereverberation model, it is mandatory to address the noise first to prevent large performance losses. If the dereverberation model is applied first, artefacts are introduced that the denoising network cannot recover from.

For the task of dereverberation, a more pronounced behaviour is visible. 
As shown in \Cref{tab:wsj_reverb}, even for the comparatively easier WSJ-Reverb dataset, applying the redundant denoising model first results in an increase of more than \num{70} percentage points in the WER, as too much of the reverberant speech is interpreted as noise and removed. Only if the dereverberation component is applied first, the performance is kept stable in a cascaded processing.
The same holds for the EARS-Reverb subset evaluated in \Cref{tab:ears_reverb}.
The joint system again suffers from some performance degradation in terms of WER, compared to the mono-objective model, indicating that it is more prone to inducing artefacts during enhancement. This can again be mitigated by incorporating signals containing only reverb during training. 
Despite the train-test mismatch, both models provide good enhancement in terms of ESTOI and DNSMOS, on par with the mono-objective and better \gls{SRMR} results.

In summary, for environments with a single disturbance present, it can be seen that a model trained to address both objectives still achieves good enhancement quality. 
If, however, a cascade of systems is employed, the active distortion needs to be mitigated first to prevent the system from significantly degrading. Only for denoising in less challenging environments, the stages can be changed in their order.

\begin{table}[bt]
    \caption{Performance of the mono-objective and multi-objective systems for denoising on WSJ-CHiME3 and EARS-WHAM.}
        \vspace{-0.75em}
    \label{tab:results_denoise}
    \begin{subtable}{\columnwidth}
    \caption{WSJ-CHiME3\vspace{-0.5em}}\label{tab:wsj_noise}
    \setlength{\tabcolsep}{4pt}
    \begin{tabular}{l c c c c }
    \toprule
         Objective & {SRMR} & {DNSMOS} & {ESTOI} & {WER}\\ 
         \midrule
         Obs. & \num{4.61} \footnotesize{$\pm$ \num{2.08}} & \num{3.10} \footnotesize{$\pm$ \num{0.39}} & \num{0.79} \footnotesize{$\pm$ \num{0.14}} & \num{6.44}\\
         \midrule

         Denoise & \num{6.09} \footnotesize{$\pm$ \num{2.57}} & \num{3.96} \footnotesize{$\pm$ \num{0.19}} & $\mathbf{0.93}$ \footnotesize{$\pm$ \num{0.05}} & $\mathbf{5.53}$\\
         Den.$\veryshortarrow$Der. & $\mathbf{6.17}$ \footnotesize{$\pm$ \num{2.64}} & $\mathbf{3.98}$ \footnotesize{$\pm$ \num{0.19}} & $\mathbf{0.93}$ \footnotesize{$\pm$ \num{0.05}} & \num{5.68}\\
         Der.$\veryshortarrow$Den. & $\mathbf{6.17}$ \footnotesize{$\pm$ \num{2.60}} & \num{3.95} \footnotesize{$\pm$ \num{0.20}} & $\mathbf{0.93}$ \footnotesize{$\pm$ \num{0.05}} & \num{5.58}\\
         Joint & \num{6.16} \footnotesize{$\pm$ \num{2.62}} & \num{3.79} \footnotesize{$\pm$ \num{0.23}} & \num{0.91} \footnotesize{$\pm$ \num{0.05}} & \num{6.25}\\
         Mixed & \num{6.06} \footnotesize{$\pm$ \num{2.54}} & \num{3.92} \footnotesize{$\pm$ \num{0.20}} & \num{0.92} \footnotesize{$\pm$ \num{0.05}} & \num{5.90}\\
         
        \bottomrule
    \end{tabular}
\end{subtable}

\vspace{1em}
\begin{subtable}{\columnwidth}
    \centering
    \setlength{\tabcolsep}{3.75pt}
    \caption{EARS-WHAM\vspace{-0.5em}}\label{tab:ears_noise}

    \begin{tabular}{l c c c c }
    \toprule
         Objective & {SRMR} & {DNSMOS} & {ESTOI} & {WER}\\ 
         \midrule
         Obs. & \num{4.50} \footnotesize{$\pm$ \num{2.54}} & \num{2.72} \footnotesize{$\pm$ \num{0.30}} & \num{0.49} \footnotesize{$\pm$ \num{0.16}} & \num{32.38}\\
         \midrule
         Denoise & $\mathbf{8.24}$ \footnotesize{$\pm$ \num{3.94}} & $\mathbf{3.88}$ \footnotesize{$\pm$ \num{0.27}} & $\mathbf{0.73}$ \footnotesize{$\pm$ \num{0.14}} & $\mathbf{16.00}$\\
         Den.$\veryshortarrow$Der. & \num{8.15} \footnotesize{$\pm$ \num{3.07}} & \num{3.73} \footnotesize{$\pm$ \num{0.31}} & \num{0.69} \footnotesize{$\pm$ \num{0.12}} & \num{17.21}\\
         Der.$\veryshortarrow$Den. & \num{7.34} \footnotesize{$\pm$ \num{3.23}} & \num{3.56} \footnotesize{$\pm$ \num{0.37}} & \num{0.63} \footnotesize{$\pm$ \num{0.16}} & \num{26.05}\\
         Joint & \num{8.17} \footnotesize{$\pm$ \num{3.87}} & \num{3.71} \footnotesize{$\pm$ \num{0.34}} & \num{0.63} \footnotesize{$\pm$ \num{0.16}} & \num{27.91}\\
         Mixed & \num{8.03} \footnotesize{$\pm$ \num{3.77}} & $\mathbf{3.88}$ \footnotesize{$\pm$ \num{0.27}} & \num{0.70} \footnotesize{$\pm$ \num{0.15}} & \num{20.23}\\
         
        \bottomrule
    \end{tabular}
\end{subtable}

\end{table}

\begin{table}[bt]
    \caption{Performance of the mono-objective and multi-objective systems for dereverberation on WSJ-Reverb and EARS-Reverb.} 
    \label{tab:results_dereverb}
    \setlength{\tabcolsep}{3.25pt}
    \vspace{-0.75em}
\begin{subtable}{\columnwidth}
    \caption{WSJ-Reverb\vspace{-0.5em}} \label{tab:wsj_reverb}
    \begin{tabular}{l c c c c }
    \toprule
         Objective & {SRMR} & {DNSMOS} & {ESTOI} & {WER}\\ 
         \midrule
         Obs. & \num{2.74} \footnotesize{$\pm$ \num{0.89}} & \num{3.04} \footnotesize{$\pm$ \num{0.3}} & \num{0.44} \footnotesize{$\pm$ \num{0.13}} & \num{19.18}\\
         \midrule
         Dereverb & \num{6.38} \footnotesize{$\pm$ \num{2.69}} & \num{4.01} \footnotesize{$\pm$ \num{0.19}} & $\mathbf{0.90}$ \footnotesize{$\pm$ \num{0.05}} & \hphantom{7}$\mathbf{6.33}$\\
         Den.$\veryshortarrow$Der. & $\mathbf{7.34}$ \footnotesize{$\pm$ \num{3.84}} & \num{3.60} \footnotesize{$\pm$ \num{0.38}} & \num{0.40} \footnotesize{$\pm$ \num{0.21}} & \num{78.24}\\
         Der.$\veryshortarrow$Den. & \num{6.39} \footnotesize{$\pm$ \num{2.70}} & $\mathbf{4.03}$ \footnotesize{$\pm$ \num{0.19}} & $\mathbf{0.90}$ \footnotesize{$\pm$ \num{0.05}} & \hphantom{7}\num{6.36}\\
         Joint & \num{6.50} \footnotesize{$\pm$ \num{2.81}} & \num{4.01} \footnotesize{$\pm$ \num{0.20}} & \num{0.87} \footnotesize{$\pm$ \num{0.06}} & \hphantom{7}\num{7.77}\\
         Mixed & \num{6.33} \footnotesize{$\pm$ \num{2.71}} & \num{4.01} \footnotesize{$\pm$ \num{0.19}} & \num{0.87} \footnotesize{$\pm$ \num{0.05}} & \hphantom{7}\num{7.86}\\
         
        \bottomrule
    \end{tabular}
    \end{subtable}
    
\vspace{1em}
\begin{subtable}{\columnwidth}
    \setlength{\tabcolsep}{2.25pt}
    \caption{EARS-Reverb\vspace{-0.5em}}\label{tab:ears_reverb}
    \begin{tabular}{l c c c c }
    \toprule
         Objective & {SRMR} & {DNSMOS} & {ESTOI} & {WER}\\ 
         \midrule
         Obs. & \hphantom{1}\num{3.98} \footnotesize{$\pm$ \num{2.31}} & \num{3.11} \footnotesize{$\pm$ \num{0.34}} & \num{0.48} \footnotesize{$\pm$ \num{0.16}} & \num{29.32}\\
         \midrule
         Dereverb & \hphantom{1}\num{7.84} \footnotesize{$\pm$ \num{3.77}} & \num{3.88} \footnotesize{$\pm$ \num{0.27}} & $\mathbf{0.86}$ \footnotesize{$\pm$ \num{0.10}} & \hphantom{7}$\mathbf{9.06}$\\
         Den.$\veryshortarrow$Der. & $\mathbf{15.11}$ \footnotesize{$\pm$ \num{11.87}} & \num{2.86} \footnotesize{$\pm$ \num{0.56}} & \num{0.35} \footnotesize{$\pm$ \num{0.22}} & \num{79.94}\\
         Der.$\veryshortarrow$Den. & \hphantom{1}\num{7.85} \footnotesize{$\pm$ \num{3.77}} & \num{3.88} \footnotesize{$\pm$ \num{0.27}} & \num{0.85} \footnotesize{$\pm$ \num{0.10}} & \hphantom{7}\num{9.20}\\
         Joint & \hphantom{1}\num{8.92} \footnotesize{$\pm$ \num{4.32}} & $\mathbf{3.89}$ \footnotesize{$\pm$ \num{0.28}} & \num{0.81} \footnotesize{$\pm$ \num{0.11}} & \num{11.69}\\
         Mixed & \hphantom{1}\num{8.10} \footnotesize{$\pm$ \num{3.87}} & \num{3.88} \footnotesize{$\pm$ \num{0.28}} & \num{0.84} \footnotesize{$\pm$ \num{0.10}} & \hphantom{7}\num{9.73}\\
        \bottomrule
    \end{tabular}
    \end{subtable}

\end{table}

\vspace{-0.15em}
\subsection{Joint denoising and dereverberation}

\subsubsection{Results on real recordings in high-SNR environments}
\vspace{-0.15em}
For the low-noise, i.e., high-SNR scenario, the models trained on WSJ data and the derived disturbed variants are  evaluated on the real test data of the REVERB challenge. This data contains stationary ambient noise, whose exact \gls{SNR} is unknown. However, the SNR for the simulated data of the challenge was set to \SI{20}{\decibel} and real data is supposed to have similar characteristics \cite{reverb_13}.

The results can be seen in the left column of \Cref{tab:joint_results}.
Interestingly, the dereverberation model achieves the best WER of all models.  The \gls{ASR} model seems to be able to handle the residual noise, but the corresponding perceptive DNSMOS metric and the SRMR are the lowest.
The cascade of the denoising and dereverberation models fails to enhance the signal, while the cascade of the dereverberation and the denoising model is able to provide a solid signal enhancement, confirming the conclusion of the previous section, that the dominating distortion type should be tackled first.

The enhancement of the two models trained in the presence of both disturbances achieves decent results w.r.t. all performance metrics.
However, note that these models were trained with an equal amount of training data compared to the single task, and the inference time is half compared to the  model cascades.


\subsubsection{Results in low-SNR environments}
The EARS-WHAMR! dataset is characterized by low \gls{SNR} and high reverberation conditions. The results on these challenging conditions are shown in the right column of \Cref{tab:joint_results}. The two models tuned to the individual distortion types and the two model cascades fail in this scenario. It can be hypothesized that the cascade of the dereverberation model followed by the denoising model, which previously worked in the high-SNR scenario, is no longer able to remove a sufficient amount of reverberation in the first stage to ease the task  in the second stage. Also, the dereverberation model, if applied as the sole model, is no longer able to achieve good WER results, as the remaining noise cannot be compensated by the \gls{ASR} model.
The joint and the mixed models perform the best of all enhancement approaches and provide a significant improvement compared to the noisy and reverberant input signals.
However, considering the challenging dataset, artifacts are still introduced by the joint and mixed model, especially in negative SNR conditions that still lead to a high WER. 
We provide example audio files for listening tests.\footnote{\url{https://go.upb.de/itg25_diffusion}}

\section{Conclusions}
In this work, we investigated the behavior of diffusion-based speech enhancement models in  reverberant, noisy, and noisy reverberant environments. The results showed that, not surprisingly, models tuned to the specific type of distortion perform best if that particular distortion type is solely present.  For the practically more relevant case of data that is both degraded by reverberation and noise, a cascade of models tuned to one of the distortions each, leads to decent results, if applied in the order of the dominating distortion. However, the extra effort incurred by two models applied sequentially can be saved by employing a single model trained on both types of distortions. Here, the preferred choice is the \enquote{Mixed Objective} model that is trained on all three types of data with equal shares. It achieves decent performance on all test conditions w.r.t. both signal enhancement and WER metrics, thus being a ubiquitous approach, that can be safely applied to unknown conditions.


\newpage
\balance
\small
\bibliographystyle{ieeetr}
\bibliography{example}

\begin{thebibliography}{10}

\bibitem{23_ochieng_se_overview}
P.~Ochieng, ``Deep neural network techniques for monaural speech enhancement
  and separation: state of the art analysis,'' {\em Artificial Intelligence
  Review}, vol.~56, no.~Suppl 3, pp.~3651--3703, 2023.

\bibitem{24_haeb_se_overview}
R.~Haeb-Umbach, T.~Nakatani, M.~Delcroix, C.~Boeddeker, and T.~Ochiai,
  ``Microphone array signal processing and deep learning for speech
  enhancement: Combining model-based and data-driven approaches to parameter
  estimation and filtering,'' {\em IEEE Signal Processing Magazine}, vol.~41,
  no.~6, pp.~12--23, 2024.

\bibitem{22_Lu_diffusion_se}
Y.-J. Lu, Z.~Wang, S.~Watanabe, A.~Richard, C.~Yu, and Y.~Tsao, ``Conditional
  diffusion probabilistic model for speech enhancement,'' in {\em IEEE
  International Conference on Acoustics, Speech and Signal Processing
  (ICASSP)}, pp.~7402--7406, 2022.

\bibitem{welker_22_sgmse}
S.~Welker, J.~Richter, and T.~Gerkmann, ``Speech enhancement with score-based
  generative models in the complex stft domain,'' in {\em Proceedings ISCA
  Interspeech}, pp.~2928--2932, 2022.

\bibitem{23_Gonzalez_heun_binaural_se}
P.~Gonzalez, Z.-H. Tan, J.~{\O}stergaard, J.~Jensen, T.~S. Alstr{\o}m, and
  T.~May, ``Diffusion-based speech enhancement in matched and mismatched
  conditions using a heun-based sampler,'' in {\em IEEE International
  Conference on Acoustics, Speech and Signal Processing (ICASSP)},
  pp.~10431--10435, 2023.

\bibitem{24_Jukic_schroedinger_diff}
A.~Juki\'{c}, R.~Korostik, J.~Balam, and B.~Ginsburg, ``Schr{\"o}dinger bridge
  for generative speech enhancement,'' in {\em Proceedings ISCA Interspeech},
  pp.~1175--1179, 2024.

\bibitem{lemercier_23_restoration}
J.-M. Lemercier, J.~Richter, S.~Welker, and T.~Gerkmann, ``Analysing
  diffusion-based generative approaches versus discriminative approaches for
  speech restoration,'' in {\em IEEE International Conference on Acoustics,
  Speech and Signal Processing (ICASSP)}, pp.~1--5, 2023.

\bibitem{lemercier_25_buddy}
J.-M. Lemercier, E.~Moliner, S.~Welker, V.~V{\"a}lim{\"a}ki, and T.~Gerkmann,
  ``Unsupervised blind joint dereverberation and room acoustics estimation with
  diffusion models,'' {\em arXiv preprint arXiv:2408.07472}, 2024.

\bibitem{22_Lu_diffusion_enhancement}
Y.-J. Lu, Z.-Q. Wang, S.~Watanabe, A.~Richard, C.~Yu, and Y.~Tsao,
  ``Conditional diffusion probabilistic model for speech enhancement,'' in {\em
  IEEE International Conference on Acoustics, Speech and Signal Processing
  (ICASSP)}, pp.~7402--7406, 2022.

\bibitem{richter_23_sgmse+}
J.~Richter, S.~Welker, J.-M. Lemercier, B.~Lay, and T.~Gerkmann, ``Speech
  enhancement and dereverberation with diffusion-based generative models,''
  {\em IEEE/ACM Transactions on Audio, Speech, and Language Processing},
  vol.~31, pp.~2351--2364, 2023.

\bibitem{lemercier_23_storm}
J.-M. Lemercier, J.~Richter, S.~Welker, and T.~Gerkmann, ``Storm: A
  diffusion-based stochastic regeneration model for speech enhancement and
  dereverberation,'' {\em IEEE/ACM Transactions on Audio, Speech, and Language
  Processing}, vol.~31, pp.~2724--2737, 2023.

\bibitem{scheibler24_interspeech}
R.~Scheibler, Y.~Fujita, Y.~Shirahata, and T.~Komatsu, ``Universal score-based
  speech enhancement with high content preservation,'' in {\em Proceedings ISCA
  Interspeech}, pp.~1165--1169, 2024.

\bibitem{lemercier_23_posterior_sampling}
J.-M. Lemercier, S.~Welker, and T.~Gerkmann, ``Diffusion posterior sampling for
  informed single-channel dereverberation,'' in {\em Proc. IEEE WASPAA},
  pp.~1--5, 2023.

\bibitem{urgent_challenge}
K.~Saijo, W.~Zhang, S.~Cornell, R.~Scheibler, C.~Li, Z.~Ni, A.~Kumar, M.~Sach,
  Y.~Fu, W.~Wang, T.~Fingscheidt, and S.~Watanabe, ``Interspeech 2025 urgent
  speech enhancement challenge,'' 2025.

\bibitem{kimura_24_msgmse}
R.~Kimura, T.~Nakatani, N.~Kamo, D.~Marc, S.~Araki, T.~Ueda, and S.~Makino,
  ``Diffusion model-based mimo speech denoising and dereverberation,'' in {\em
  International Conference on Acoustics, Speech, and Signal Processing
  Workshops (ICASSPW)}, pp.~455--459, 2024.

\bibitem{nakatani_24_msgmse_multistream}
T.~Nakatani, N.~Kamo, M.~Delcroix, and S.~Araki, ``Multi-stream diffusion model
  for probabilistic integration of model-based and data-driven speech
  enhancement,'' in {\em Proc. IEEE IWAENC}, pp.~65--69, 2024.

\bibitem{23_Yang_Usee_multimodal}
M.~Yang, C.~Zhang, Y.~Xu, Z.~Xu, H.~Wang, B.~Raj, and D.~Yu, ``usee: Unified
  speech enhancement and editing with conditional diffusion models,'' {\em IEEE
  International Conference on Acoustics, Speech and Signal Processing
  (ICASSP)}, pp.~7125--7129, 2023.

\bibitem{richter_24_ears}
J.~Richter, Y.-C. Wu, S.~Krenn, S.~Welker, B.~Lay, S.~Watanabe, A.~Richard, and
  T.~Gerkmann, ``Ears: An anechoic fullband speech dataset benchmarked for
  speech enhancement and dereverberation,'' in {\em Proceedings ISCA
  Interspeech}, pp.~4873--4877, 2024.

\bibitem{reverb_13}
K.~Kinoshita, M.~Delcroix, T.~Yoshioka, T.~Nakatani, E.~Habets, R.~Haeb-Umbach,
  V.~Leutnant, A.~Sehr, W.~Kellermann, R.~Maas, S.~Gannot, and B.~Raj, ``The
  {REVERB} challenge: A common evaluation framework for dereverberation and
  recognition of reverberant speech,'' in {\em Proc. IEEE WASPAA}, pp.~1--4,
  2013.

\bibitem{vincent_11_denoising_score_matching}
P.~Vincent, ``A connection between score matching and denoising autoencoders,''
  {\em Neural Computation}, vol.~23, no.~7, pp.~1661--1674, 2011.

\bibitem{song_21_sgm}
Y.~Song, J.~Sohl-Dickstein, D.~P. Kingma, A.~Kumar, S.~Ermon, and B.~Poole,
  ``Score-based generative modeling through stochastic differential
  equations,'' in {\em International Conference on Learning Representations},
  2021.

\bibitem{sgmse_repo}
{Signal Processing (SP), Universit\"at Hamburg}, ``{Speech Enhancement and
  Dereverberation with Diffusion-based Generative Models}.''
  \url{https://github.com/sp-uhh/sgmse/}, 2025.

\bibitem{wsj}
J.~Garofolo, D.~Graff, D.~Paul, and D.~Pallett, ``{CSR-I (WSJ0) Complete
  LDC93S6A},''
\newblock {Web Download, Philadelphia: Linguistic Data Consortium, 1993}.

\bibitem{dnsmos_21}
C.~K.~A. Reddy, V.~Gopal, and R.~Cutler, ``Dnsmos: A non-intrusive perceptual
  objective speech quality metric to evaluate noise suppressors,'' in {\em IEEE
  International Conference on Acoustics, Speech and Signal Processing
  (ICASSP)}, pp.~6493--6497, 2021.

\bibitem{srmr}
T.~H. Falk, C.~Zheng, and W.-Y. Chan, ``A non-intrusive quality and
  intelligibility measure of reverberant and dereverberated speech,'' {\em IEEE
  Transactions on Audio, Speech, and Language Processing}, vol.~18, no.~7,
  pp.~1766--1774, 2010.

\bibitem{estoi}
J.~Jensen and C.~H. Taal, ``An algorithm for predicting the intelligibility of
  speech masked by modulated noise maskers,'' {\em IEEE/ACM Transactions on
  Audio, Speech, and Language Processing}, vol.~24, no.~11, pp.~2009--2022,
  2016.

\bibitem{roux_19_sisdr}
J.~L. Roux, S.~Wisdom, H.~Erdogan, and J.~R. Hershey, ``{SDR - Half-baked or
  Well Done?},'' in {\em IEEE International Conference on Acoustics, Speech and
  Signal Processing (ICASSP)}, pp.~626--630, 2019.

\bibitem{drude2019sms}
L.~Drude, J.~Heitkaemper, C.~Boeddeker, and R.~Haeb-Umbach, ``Sms-wsj:
  Database, performance measures, and baseline recipe for multi-channel source
  separation and recognition,'' {\em arXiv preprint arXiv:1910.13934}, 2019.

\bibitem{nemo}
O.~Kuchaiev, J.~Li, H.~Nguyen, O.~Hrinchuk, R.~Leary, B.~Ginsburg, S.~Kriman,
  S.~Beliaev, V.~Lavrukhin, J.~Cook, P.~Castonguay, M.~Popova, J.~Huang, and
  J.~M. Cohen, ``Nemo: a toolkit for building {AI} applications using neural
  modules,'' {\em CoRR}, vol.~abs/1909.09577, 2019.

\end{thebibliography}


\end{document}